\documentclass[10pt]{iopart}
\usepackage{graphicx,epsf}

\usepackage[utf8]{inputenc}
\usepackage{calc}
\usepackage{dcolumn}

\begin{document}

\MSMSE

\title{Density Functional Theory investigations of titanium $\gamma$-surfaces and stacking faults}
\author{Magali Benoit}
\address{CEMES-CNRS UPR 8011, 29 rue Jeanne Marvig, 31055 Toulouse Cedex, France }
\author{Nathalie Tarrat}
\address{Present adress: Université de Toulouse; INSA,UPS,INP; LISBP, 135 Avenue de Rangueil, F-31077 Toulouse, France;  INRA, UMR792, Ingénierie des Systèmes Biologiques et des Procédés, F-31400 Toulouse, France;
CNRS, UMR5504, F-31400 Toulouse, France}
\author{Joseph Morillo}
\address{CEMES-CNRS UPR 8011, 29 rue Jeanne Marvig, 31055 Toulouse Cedex, France \\
Universit\'e Paul Sabatier, 118 route de Narbonne, F-31062 Toulouse Cedex 9, France }

\begin{abstract}

Properties of hcp-Ti such as elastic constants, stacking faults and $\gamma {\rm -surfaces}$ are  computed using Density Functional Theory (DFT) and
two central force Embedded Atom interaction Models (EAM) \cite{Zope2003,Hammerschmidt2005}.
The results are compared to previously published calculations and to predicting models.
Their implications on the plastic properties of hcp-Ti are discussed.

\end{abstract}



\maketitle

\section{Introduction}\label{sec:intro}
The plastic behavior of hexagonal compact metals (about twenty) is dominated by the movement of dislocations with the shortest Burgers vector: $\frac{a}{3}\langle1 1 \bar{2} 0\rangle$ (see D. Caillard and J.-L. Martin \cite{Caillard2003}, and M. H. Yoo and coworkers \cite{Yoo2002} for recent reviews).
For example, most of the transition metals have a basal slip plane, but  three of them (Zr, Ti and Hf) and a number of rare-earth metals (Gd, Ru, Tb, Hf, Dy and Er)  present a dominant prismatic slip plane \cite{Yoo2002,Hirth1992,Bacon2002,Legrand1984a}.
The intrinsic characteristics of dislocations (structure, stability, mobility, formation, multiplication) as well as their extrinsic characteristics (interaction with impurities, point defects, grain boundaries...)  are therefore essential ingredients for a good theoretical description of the plastic behaviour of hexagonal close packed (hcp) metals.

The core structure of the $\frac{a}{3}\langle1 1 \bar{2} 0\rangle$ screw dislocation in hcp metals has been subjected  to only a few theoretical studies (see D. J. Bacon and V. Vitek for a recent review \cite{Bacon2002}).
B. Legrand,  who introduced an explicit treatment of the electronic degrees of freedom in a tight-binding (TB) approach \cite{Legrand1984a,Legrand1985}, was the first one to obtain a prismatic spreading in an hcp metal (Ti). Moreover, this prismatic spreading was  energetically more favorable than the basal one by 24 meV/{\AA}.
He argued that this prismatic preference was related to an increased basal stacking fault energy of transition metals with $d$ fillings between 1.5 and 2.5 electrons. This effect results from the directional $d$-covalent bonding of partially filled $d$ bands, which cannot be described with pair potentials nor with more sophisticated central force N-body potentials.
More precisely, he established \cite{Legrand1984a,Legrand1984,Legrand1986} on the basis of his TB calculations (and pseudopotential calculations for divalent and trivalent metals), a clear correlation between  the basal or prismatic slip in hcp metals and the ratio $R = ( \gamma_{{\rm b}}/{\rm  C_{44}})/( \gamma_{{\rm p}}/{\rm  C_{66}})$.
$\gamma_{{\rm b}}$ and $\gamma_{{\rm p}}$ are respectively the basal $I_2$ and   the prismatic stacking faults excess energies and C$_{44}$ and  C$_{66}$ the shear elastic constants governing the shear deformation leading to the corresponding stacking faults.
This ratio measures the relative facility to form both stacking faults and thus the easiness of dislocation dissociation in the corresponding planes.
For $R > 1$, the observed slip should be prismatic whereas it should be basal for $R < 1$.
One must quote that, in all his calculations, he never found a stable prismatic stacking fault, but the $\gamma_{{\rm p}}$-surface presented a deep valley  in the $\langle \bar1 2 \bar1 0\rangle$ direction.
He defined then the stacking fault excess energy as an average value of the $\gamma_{{\rm p}}$ surface energy in this direction, corresponding to a continuous dislocation spreading.
Notwithstanding, in principle, their unability to describe the directional character of the bonding in Ti, pair potentials and N-body central force potentials have been able to reproduce the preferential prismatic core structure of the $\frac{a}{3}\langle1 1 \bar{2} 0\rangle$ screw dislocation in some cases \cite{Liang1986,Vitek1991}.

More recently, A. Girshick and coworkers \cite{Girshick1998a} derived a bond order (BO) potential for Ti \cite{Girshick1998} allowing them to
deduce more accurate values of the excess energies of the basal and prismatic dislocation cores.
With the BO potential, in agreement with B. Legrand \cite{Legrand1984a,Legrand1985}, they found that the prismatic core was the most stable one (by about 20 meV/{\AA} relative to the basal one).
At odds with the BO potential, the N-body Finnis-Sinclair  central force potential \cite{Ackland1992} that they used for comparison, favored the basal spreading (by about 25 meV/{\AA}).
In agreement with B. Legrand, they did not found any stable prismatic stacking fault.
Most importantly, from their BO calculations, they obtained a $R=0.4$ value, which means a preferential basal spreading in contradiction with their determination of the core structure energies, which questions the validity of  Legrand's criterion.

Up to now, DFT calculations on the $\frac{1}{3}\langle11\bar{2}0\rangle$ screw dislocation in hcp metals (Zr and Ti) were reported by C. Domain and A. Legris \cite{Domain2002,Domain2004} on very small clusters and by N. Tarrat {\it et al.} \cite{Tarrat2009} on slightly larger ones. Both studies showed a prismatic spreading with a core structure in overall agreement with the previous TB calculations \cite{Legrand1984a,Legrand1985,Girshick1998a}. Very recently,  DFT calculations on very large clusters using flexible boundary conditions have been reported \cite{Ghazisaeidi2012}. They found similar prismatic spreadings for the $\frac{1}{3}\langle11\bar{2}0\rangle$ screw dislocation than the ones obtained by Tarrat {et al.}, but argued that the prismatic core structures are slightly different and that one of them is  actually a metastable one.

The purpose of this paper is therefore to present accurate calculations of the bulk properties, stacking faults and $\gamma$-surfaces which are essential properties to gain a deeper understanding of the relationship between the electronic structure of hcp Ti and its plastic behavior.
The DFT results are compared to previously published calculations (TB \cite{Legrand1984a,Legrand1985}, BO potential \cite{Girshick1998a}. MEAM potential \cite{Hennig2008} and DFT calculations \cite{Domain2002,Domain2004,Hennig2008}) and to similar calculations performed with recently developed accurate semi-empirical EAM potentials \cite{Zope2003,Hammerschmidt2005}.
These EAM potentials  were specifically designed to reproduce a large data base of properties in TiAl \cite{Zope2003} or Ti \cite{Hammerschmidt2005}, either experimentally measured or deduced from \textit{ab initio} DFT calculations.
 These data bases include a large panel of configurations far from the ideal hcp lattice, so these potentials might be accurate for the description of the dislocation core structure.

 After giving the simulation details in Sec.~\protect\ref{details}, we present the results obtained on the bulk properties, stacking faults and $\gamma$-surfaces of hcp Ti using two EAM potentials and DFT calculations in Secs.~\protect\ref{ssec:res-bulk},  \protect\ref{ssec:res-faults} and \protect\ref{ssec:res-gammasurf}.
The obtained results are discussed and compared to the literature. In particular their implication on the plastic behavior of hcp Ti is examined.

\section{Simulation details}
\label{details}
\subsection{DFT calculations}
\label{ssec:simdet-DFT}
The  DFT calculations have been performed with the SIESTA \cite{Soler2002} code in which the orbitals are developed on a local basis set.
The PBE-GGA gradient approximation \cite{Perdew1996a} has been used for the exchange and correlation functional and the pseudopotential was a Trouiller-Martins \cite{Troullier1991} one,  with the 3\textit{p}4\textit{s}3\textit{d} states as valence states.
A polarized double-$\zeta$ basis set was employed for the 4$s$ electrons and a single-$\zeta$ basis set was employed for the 3$p$ and 4$d$ electrons.

A 500 Ry real space grid cutoff and a Methfessel-Paxton smearing of order one  with an electronic temperature of 300 meV were used in all calculations.
For the bulk calculations, performed with the conventional primitive cell, a  11$\times$11$\times$7 k-point mesh was used for the first Brillouin zone sampling, which was extended to a 16$\times$16$\times$12 mesh for the determination of some of the elastic constants.
For the other calculations, the specific k-point-mesh will be given in the corresponding section.

\subsection{Semiempirical potential calculations}
\label{ssec:simdet-eam}
The two semiempirical potentials  were of the embedded atom method \cite{Daw1983} type.
The first one (referred here after as ZM) was developed by R. R. Zope and Y. Mishin \cite{Zope2003} for the TiAl alloy.
They fitted  their Ti potential to experimental data (lattice parameters: $a$, $c/a$, cohesive energy $E_c$ and the five independant elastic constants C$_{ij}$) and to the \textit{ab initio} DFT volume-pressure curves of various crystal structures (hcp, face-centered cubic (fcc), body-centered cubic (bcc), simple cubic (sc) and $\omega$).
Its cutoff distance is in between the fourth and fifth neighbour shells.
The second one (referred here after as HKV)  was developed by  Hammerschmidt and coworkers \cite{Hammerschmidt2005} for the description of grain boundaries.
It was also fitted to bulk Ti properties and to a large data base of low coordinated configurations, namely small clusters (up to 8 atoms), surfaces  and adatom surface diffusion.
Its cutoff distance is in between the third and fourth neighbour shells.
Both of them are then expected to be able to describe configurations far from the ideal hcp structure.

In all calculations where periodic conditions were applied in a given direction, the simulation cell dimension in that direction was set to at least five times the interaction cutoff  in order to avoid any spurious interaction of an atom with its periodic images.

\section{Results}
\label{sec:results}

\subsection{Bulk properties}
\label{ssec:res-bulk}

In Tables~\protect\ref{tab:props} and \ref{tab:elasticconst}, the bulk properties of hcp Ti calculated with DFT and the two EAM potentials are compared to previous calculations performed with DFT \cite{Hennig2008}, the BO potential \cite{Girshick1998} and a new MEAM potential \cite{Hennig2008} and to experimental data.
\begin{table}[ht]
\caption{\label{tab:props} Properties of bulk hcp Ti predicted by the DFT calculations and the two EAM potentials \cite{Zope2003,Hammerschmidt2005}  compared to previously published   calculations \cite{Girshick1998,Hennig2008} and to experimental data \cite{Fisher1964}.
Lattice parameter  $a$ is in {\AA}, $V_0$ is in {\AA}$^3$/at., the bulk modulus $B$ is in GPa.}
\begin{center}
\begin{tabular}{lllll}
\hline\hline
			    & a 	& c/a 	& V$_0$ 	& B	 \\
\hline
DFT (this work)       	    & 2.996  & 1.588 & 18.49 & 110.2  	\\
DFT\cite{Hennig2008} 	    & 2.947 &  1.583 & 17.55 & - \\
ZM (this work)     	    & 2.951  & 1.585 & 17.64 & 110.5          \\
HKV (this work) & 2.969  & 1.590 & 18.02 & 110.4          \\
BO\cite{Girshick1998} 	    & 2.950  & 1.587 & 17.65 & 113.6          \\
MEAM\cite{Hennig2008} 	    & 2.930  & 1.596 & 17.40 & - 					\\
\hline
Exp.\cite{Fisher1964} 	    & 2.951  & 1.588 & 17.65 & 110            \\
\hline\hline
\end{tabular}
\end{center}
 \end{table}

The equilibrium volume and bulk modulus are obtained with a similar accuracy in all cases.
For the elastic constants, the agreement between the DFT calculations and experiment is very satisfactory (deviations less than 10\%), better than with the EAM, BO or MEAM potentials (deviations up to $\simeq$ 40\%, 20\% and 15 \% for the EAM, the BO and the MEAM potentials, respectively).
Notably, there is a large deviation on C$_{66}$ with the EAM potentials.
It is particularly interesting to note that the Cauchy pressures, which are due to N-body interactions only, are rather well reproduced by the present DFT calculations and the MEAM ones, whereas the deviations from experimental values are large (up to 90\%) using the EAM or the BO potentials for which  C$_{13}$ - C$_{44}$  is overestimated, whereas  C$_{12}$ - C$_{66}$ is underestimated.
The MEAM potential appears then as the best interaction model for the description of the Ti elastic properties.
The small differences between the present DFT calculations and the ones of Ref.\cite{Hennig2008} are certainly due to the use of a different GGA functional and a different basis set. 
 \begin{table}[ht]
\caption{\label{tab:elasticconst} Elastic constants of bulk hcp Ti predicted by the DFT calculations and the two EAM potentials \cite{Zope2003,Hammerschmidt2005}  compared to previously published  calculations \cite{Girshick1998,Hennig2008} and to experimental data \cite{Fisher1964}.
All elastic constants C$_{ij}$ and Cauchy pressures (CP$_1$=C$_{12}$-C$_{66}$ and CP$_2$=C$_{13}$-C$_{44}$) are in GPa.
 Note that the bond order potential of Ref. \cite{Girshick1998}
was fitted exactly on the C$_{11}$, C$_{33}$ and C$_{44}$ elastic constants.}
\begin{center}
 \begin{tabular}{lllllllll}
\hline\hline
			&  C$_{11}$  & C$_{12}$ & C$_{13}$ & C$_{33}$ & C$_{44}$  & C$_{66}$ & CP$_1$ & CP$_2$ \\
\hline
DFT (this work) 	&  183.4  & 84.6 	 & 63.8 	 & 204.9 	& 48.8	 & 49.4  & 35.2	 & 15.0	 \\
DFT\cite{Hennig2008} 	& 172     &   82         &  75          &  190          &  45 &  45   &  37   & 30  \\
ZM (this work)    	    &  186.2 &  69.5  &  76.2  & 189.4  & 46.4  &   58.3  & 11.2  & 29.8  \\
HKV (this work) & 188.6  & 65.4   &  67.4  &  216.9 & 45.8  &   61.6  & 3.8   & 21.6 \\
BO\cite{Girshick1998} 	    & 176.0   & 74.0  &  83.3  &   190.5& 50.8   & 51.1    &  22.9  & 32.5  \\
MEAM\cite{Hennig2008} 	    & 174     &  95   &  72    &  188   &  58    &  39.5   & 55.5  &  14 \\
\hline
Exp.\cite{Fisher1964} 	    & 176.1   &  86.9 & 68.3  &  190.5 & 50.8  &  44.6  &  42.3  &  17.5   \\
\hline\hline

\end{tabular}
\end{center}
\end{table}

\subsection{Stacking faults}
\label{ssec:res-faults}
The excess energies of the two intrinsic I$_1$ (ABAB$|$CBCB) and I$_2$ (ABAB$|$CACA), and the extrinsic I$_E$ (ABAB$|$C$|$ABAB) elementary stacking faults in the basal plane have been determined,
as well as the easy prismatic stacking fault.
They have been calculated in a slab geometry with free surfaces.

The excess fault energy $\gamma$ is then given by:
\begin{equation}
 \gamma = E_{{\rm faulted}}-E_{{\rm perfect}},
\end{equation}

where $E_{{\rm faulted}}$ is the total energy of the slab containing the fault, and $E_{{\rm perfect}}$ is the total energy of the perfect slab.
For the DFT calculations, the size of the vacuum (4 interatomic distances, i.e. $\approx$ 9.5 {\AA}) and the number of planes (14 atomic planes for I$_1$ and I$_2$ and 15 atomic planes for I$_E$) have been adjusted to have good converged results.
The slabs were then relaxed, using a mesh of 12x12x2 k-points.
In the EAM simulations, the slabs were made of 40 atomic planes and were replicated 8 times in the fault plane, so that the dimensions of the supercell are  about 5 times the cutoff radius of the potential. 

\begin{figure}[ht]
\begin{center}
 \includegraphics[scale=0.3]{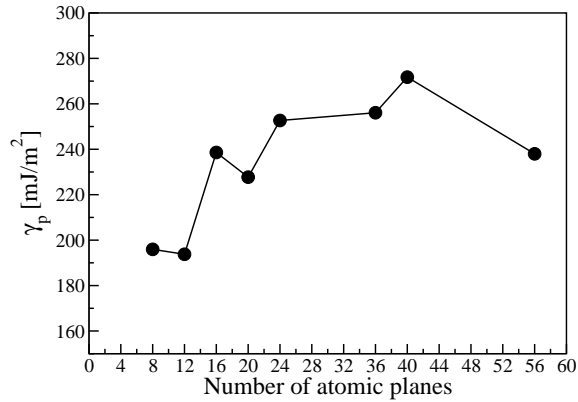}
\caption{Evolution of the DFT prismatic stacking fault excess energy as a function of the number of atomic planes. }
\end{center}
\label{fig:prism}
\end{figure}

The stacking fault energies, together with the (0001) surface energy, are presented in Table~\protect\ref{tab:faults} and compared to those obtained with the EAM potentials, to experiments and to previously published results \cite{Legrand1984a,Legrand1985,Girshick1998a,Domain2002,Domain2004,Hennig2008,Ghazisaeidi2012}.
\begin{table}[ht]

\caption{\label{tab:faults} Hcp Ti  (0001) surface energy $\gamma_s$ and stacking fault excess energies (mJ/m$^2$): $\gamma_{\rm I_1}$, $\gamma_{\rm I_2}$, $\gamma_{\rm I_E}$ (0001) stacking faults and  $\gamma_{\rm p}$ prismatic stacking fault.
Present DFT and EAM  (ZM \cite{Zope2003} and HKV \cite{Hammerschmidt2005}) potentials calculations are compared to experimental data and previous calculations with MEAM, BO, TB and DFT  methods.}
\begin{tabular}{lcccccccc}
\hline\hline
 &\multicolumn{1}{c}{$\gamma_{\rm s}$}&\multicolumn{1}{c}{$\gamma_{\rm I_1}$} &\multicolumn{1}{c}{$\gamma_{\rm I_2}$} &\multicolumn{1}{c}{$\gamma_{\rm I_E}$} &\multicolumn{1}{c}{ $\gamma_{\rm p, easy}$}\\
\hline
DFT (this work)	&\multicolumn{1}{c}{2048}	& 148.6			& 259.1		& 353.1		& 250.0			\\
ZM (this work)	&\multicolumn{1}{c}{1263}	& 30.6			& 54.3		& 82.4		& 364.6			\\
ZM\cite{Zope2003}&\multicolumn{1}{c}{1725}	& 31			& 56		& 82		& -			\\
HKV (this work)  &\multicolumn{1}{c}{1188}	& 33.6			& 64.5		& 94.2		& 353.6			\\
HKV\cite{Hammerschmidt2005}&\multicolumn{1}{c}{1185}& -			& -		& -		& -			\\
A-FS\cite{Girshick1998}	   & -			& 33			& 64		& 94		& 253			\\
BO\cite{Girshick1998}	   &-			& -			& 110		& -		& 260			\\
TB\cite{Legrand1984,Legrand1984a,Legrand1985}&-	& - 			& 290-370	& - 		& \multicolumn{1}{c}{110 -140}		\\
TB\cite{Bere1999a}	&-			& 44			& 118		& -		& -			\\
DFT\cite{Domain2002,Domain2004}	&-	 	& -			& 291		& -		& 174			\\
DFT\cite{Wu2010}& \multicolumn{1}{c}{2045}&	& 287			&-		& -		& -	\\
DFT\cite{Hennig2008,Ghazisaeidi2012}& \multicolumn{1}{c}{1938}&  -  	&  272-320 	& -      	& 220 \\
MEAM\cite{Hennig2008} & \multicolumn{1}{c}{1474}&     -     		& 170-172  	&   -        	& 297  \\
\hline
Exp.	         &\multicolumn{1}{c}{2100\cite{DeBoer1989}, 1920\cite{Tyson1977}}&  - & \multicolumn{1}{c}{$>$300\cite{Partridge1967} }	&	& \multicolumn{1}{c}{150\cite{Crecy1983}}	\\
\hline\hline
\end{tabular}
\end{table}
All the DFT calculations give very similar results expect those of  Refs.\cite{Domain2002,Domain2004} for the prismatic stacking fault value which they found 50\% lower.
This difference comes from the strong dependence of the excess energy with the number of planes in the slab.
Indeed, in Fig. \protect\ref{fig:prism}, the evolution of the prismatic stacking fault excess energy as a function of the number of atomic planes included in the calculation is presented.
The excess energy  starts to converge when the number of planes is at least greater than 16, but  with still noticeable oscillations with the number of planes.
 These oscillations are known to be due to possible quantum size effects for specific slab geometries which lead to long range interlayer oscillating relaxations \cite{jaklevic1971,jaklevic1975,schulte1976,feibelman1983,batra1986} and consequently make the DFT calculations very sensitive to the number of planes.
The present DFT value reported in Tab.~\protect\ref{tab:faults} is the mean value for calculations with a number of planes larger than 16.
Considering the very indirect experimental determination of the stacking fault excess energies, the DFT results are satisfactory.

The two EAM potentials give  identical results but those results are quite different from the DFT values in spite of the fact that they were chosen because they were developed in order to reproduce defects with high angular variations in titanium.
These differences may have important consequences for dislocation core structure calculations as well as for the evaluation of the Peierls stress.
The results obtained in the TB scheme or with the BO or MEAM potentials are much closer to the DFT ones.
This can be explained by a better reproduction of the N-body interactions.
However, surprisingly, some of those TB or BO results \cite{Legrand1984,Legrand1984a,Legrand1985} reproduce properly the basal stacking fault and poorly the prismatic one, whereas some others \cite{Girshick1998,Bere1999a} do the reverse.

The MEAM potential still appears as the best approximated interaction model, however it is still unable to reproduce the correct energetic ordering between the basal $I_2$ and the easy prismatic stacking faults which could lead to an erroneous preferential basal spreading of the $\frac{a}{3}\langle1 1 \bar{2} 0\rangle$ screw dislocation and consequently a poor description of the plastic properties. \\

\subsection{$\gamma$-surfaces}
\label{ssec:res-gammasurf}

For the determination of the $\gamma{\rm -surfaces}$, similar systems have been used, i.e. slabs with free surfaces, of 14 atomic planes for the basal surface and of 16 atomic planes for the prismatic one.
In the EAM simulations, the slabs were made of 40 planes of 64 atoms in the basal case and of 40 planes of 70 atoms in the prismatic case, leading to systems of 2560 and 2800 atoms, respectively.
\begin{figure*}[p]
\includegraphics[scale=0.95]{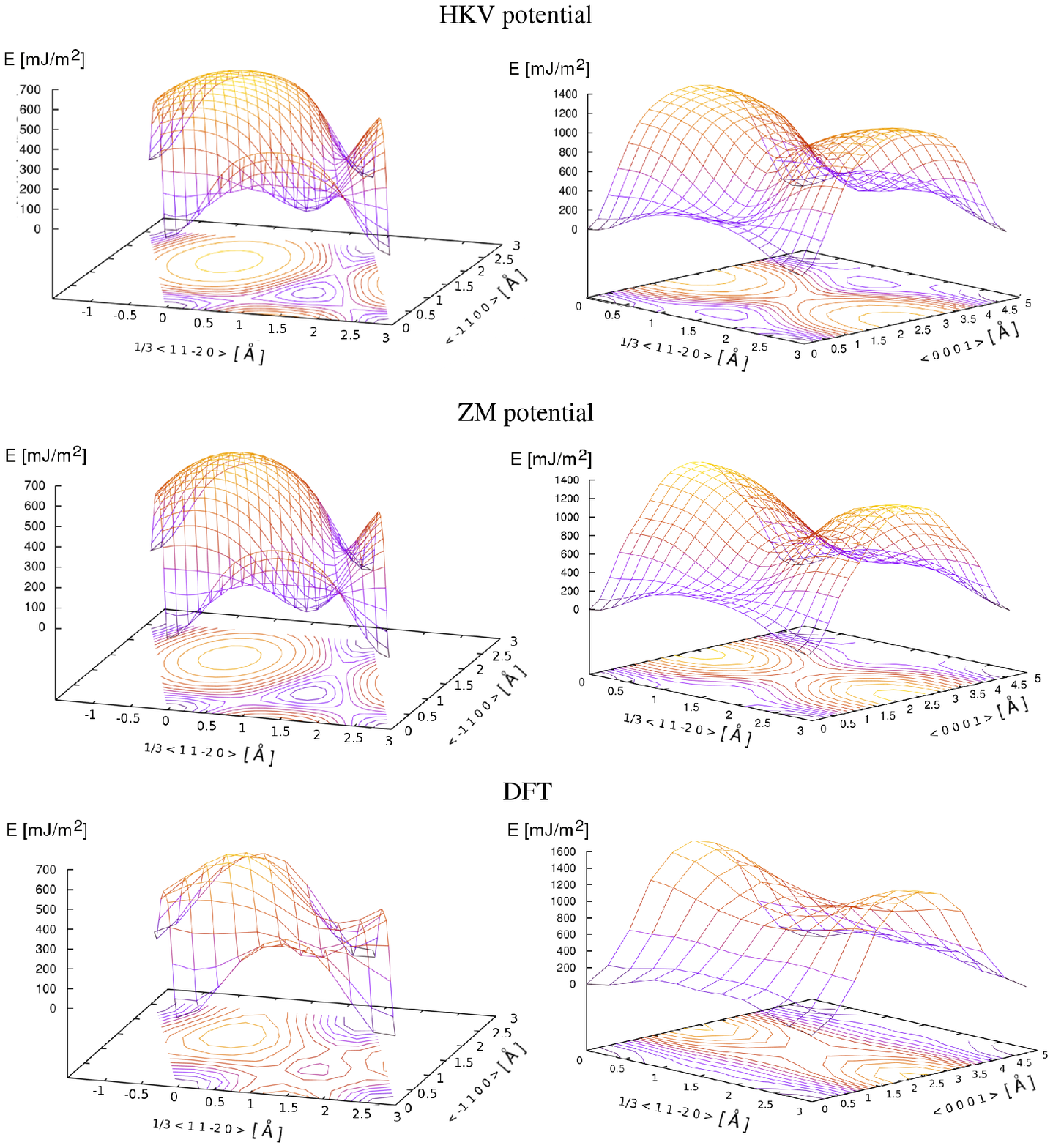}
\caption{Hcp Ti gamma surfaces. Left: basal plane, right: prismatic plane. Top and middle: HKV  and ZM  EAM potentials calculations, respectively. Bottom: DFT calculations.}
\label{fig:gsurf}
\end{figure*}

The slabs were cut into two parts, which were then shifted with respect to each other by steps of 0.005$\times$a in the empirical simulations and of 0.01$\times$a in the DFT ones.
The atoms were allowed to relax only in the direction perpendicular to the slab surface.

The basal $\gamma{\rm -surfaces}$ obtained with the two EAM potentials have a  similar global shape than the one obtained in DFT (Fig.~\protect\ref{fig:gsurf}, left).
However one can note differences in particular on the basins,  which are quite  deep using empirical potentials, and are almost inexistant using DFT.
These basins correspond to the basal stacking faults ($\gamma_{I_E}$ in Table~\protect\ref{tab:faults}), which were indeed found much lower than the DFT one.
The  only difference between the two EAM potentials appears to be a more flat maximum  in the HKV potential case than in the ZM case.
\begin{figure}[ht]
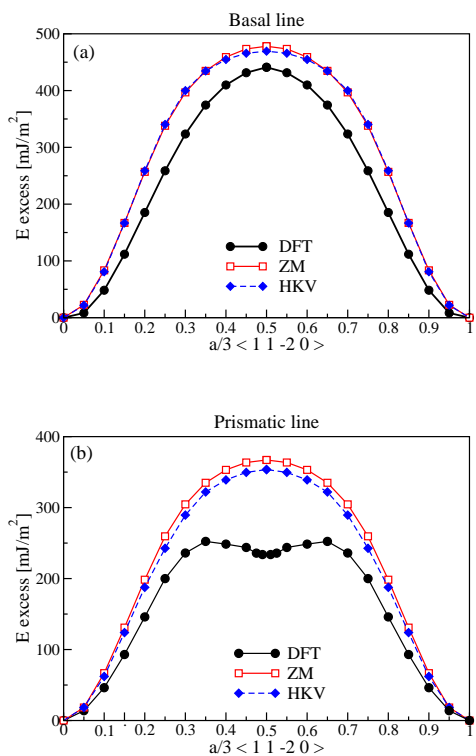

\centering
 \includegraphics[scale=0.25]{figure3a.eps} \\
\vspace{0.8cm}
\includegraphics[scale=0.25]{figure3b.eps}
\caption{Hcp Ti basal (a) and prismatic (b) $\gamma{\rm -lines}$  in the $\frac{1}{3}\langle1 1 \bar 2 0\rangle$ Burgers vector direction. Circles: DFT calculations. Squares and rhombus: ZM and HKV EAM potentials calculations, respectively.}
\label{fig:gligne_prism}
\end{figure}

The two EAM prismatic surfaces (Fig.~\protect\ref{fig:gsurf}, right) present only small differences.
The HKV surface is slightly  flatter than the ZM one.
However their differences with the DFT one are more important.
The maxima are quite lower and the basins are deeper.
One can also see that the DFT $\gamma{\rm -line}$ along a $\frac{a}{3}\langle1 1 \bar{2} 0\rangle$ direction differs notably from those of the two EAM potentials.
This result is important since it is commonly accepted that the shape of the $\gamma{\rm -line}$ is correlated with the dislocation gliding.
So an overestimation of the excess energy can lead to an error in the spreading plane.
This can be more easily seen in Fig.~\protect\ref{fig:gligne_prism} where the basal and prismatic $\gamma{\rm -lines}$ along a $\frac{a}{3}\langle1 1 \bar{2} 0\rangle$ Burgers vector  are depicted.
In the basal case (Fig.~\protect\ref{fig:gligne_prism}(a)), the differences between the DFT line and the EAM potentials ones are negligeable.
This is not true in the prismatic case (Fig.~\protect\ref{fig:gligne_prism}(b)) where the DFT line exhibits a local minimum contrary to the two EAM potentials and correspondingly its maximum is much lower. In the EAM calculations, there is also a local minimum but it is slightly shifted  from the $\frac{a}{3}\langle1 1 \bar{2} 0\rangle$ line (Fig. \protect\ref{fig:gsurf}, right).
This  local minimum of $\simeq$ 18.6 mJ/m$^2$ at the center of the line in the DFT case indicates the existence of a stable stacking fault in the prismatic plane, and thus a possible dissociation into partial dislocations.
This result was already obtained in Zr \cite{Domain2002,Domain2004} and Ti \cite{Domain2002,Domain2004,Ghazisaeidi2012} by DFT calculations but not with empirical or semi-empirical potentials \cite{Girshick1998a,Ghazisaeidi2012}.

\section{Discussion and conclusion}
\label{sec:disc}
A detailed study of bulk properties of hcp-Ti using the state of the art DFT scheme or approximated central force EAM interaction models has been presented.
Important properties for the description of dislocations have been studied: the elastic constants, the stacking faults and the $\gamma{\rm -surfaces}$.
Using the calculated stacking fault energies one can evaluate the $R$ ratio of B. Legrand \cite{Legrand1985} for the determination of the easy slip plane for the $\frac{a}{3}\langle1 1 \bar{2} 0\rangle$ screw dislocation (see the Introduction).

The DFT results give  $R=1.1$, a value slightly higher than 1, leading to a preferential prismatic spreading as experimentally observed, and in agreement with our previous DFT calculations of the $\frac{a}{3}\langle1 1 \bar{2} 0\rangle$  dislocation core structure~\cite{Tarrat2009}.
Our DFT $R$ value is consistent with the one deduced from previously published  DFT calculations (\cite{Domain2002,Domain2004,Hennig2008,Ghazisaeidi2012}) which always lead to $R$ values slightly larger than 1.
The DFT ratio, just above 1, is much less than the one  initially obtained by Legrand (2.5) in his TB calculations.
His higher value was mainly due to the underestimation of the prismatic $\gamma_{\rm p, easy}$  stacking fault energy \cite{Legrand1985} by a factor of about 2.
Unexpectedly, in its TB - BO calculations, Girschich and coworkers~\cite{Girshick1998a} found  $R=0.4$, a value much lower than 1, which is supposed to lead to a preferential basal spreading, whereas they observed a prismatic one.
In that case, surprisingly the difference comes essentially from an understimation of the basal stacking fault energy by a factor of about 2 \cite{Girshick1998a}.
A similar underestimation of this basal stacking fault was also obtained by B\'er\'e \cite{Bere1999a}.
It is surprising that these three TB calculations give so different values for the stacking fault energies, in particular for the basal one.
Indeed in the TB approach, to obtain the right energetic ordering of the hcp, fcc and bcc phases in compact close-packed structures, which conditions a good value for the basal stacking fault,  one needs to make a calculation involving up to the sixth moment of the electronic density of states for the electronic energy calculation with the used recursion method.
In Legrand's calculations six moments were used, whereas in Girshick's and B\'er\'e's calculations 9 and 32 moments were used respectively.
So, in these TB calculations the basal stacking fault excess energy, which originates mainly from the electronic part of the total energy,\footnote{The interatomic distances and the coordination numbers are unchanged up to the second neighbor shell (included), which means an essentially unchanged repulsive energy in the model.} should be described with increasing accuracy with the number of used moments.
As expected the basal stacking fault energy is rather well evaluated by Legrand, but unexpectedly this evaluation is  bad in  Girshick's and B\'er\'e's calculations with much too low values.
So, increasing the accuracy in the description of the electronic band structure, instead of improving the description of the basal stacking fault, makes it dramatically worth.
At odds, it seems to improve the description of the prismatic stacking fault excess energy.
In that case, however, since the interatomic distances are quite modified, the excess energy will not be due to the sole electronic energy.

Using the EAM results, we obtain a 0.2 $R$ value, leading to a preferential basal spreading as we do have observed in our EAM calculations of the $\frac{a}{3}\langle1 1 \bar{2} 0\rangle$  dislocation core structure~\cite{Tarrat2009}.
The introduction of angular forces in the MEAM model is supposed to allow a better description of the basal stacking fault.
Indeed, the calculated MEAM \cite{Hennig2008,Ghazisaeidi2012} basal $\gamma_{\rm I_2}$ excess energy is three times higher than the EAM one.
However, it is still lower than the DFT one leading to a $R$ ratio significantly lower than 1 (.4), since the prismatic fault energy is properly evaluated.

As previously discussed by Domain \cite{Domain2004}, the  Legrand criterion is too simplified.  In particular, it does not take into account the energy cost associated to the edge components of the dissociated basal dislocation. This extra energy cost means a preferential prismatic spreading even with $R$ values lower than one. This explains the preferential prismatic spreading observed in the DFT calculations even with an $R$ ratio only slightly larger than one, and similarly could explain  the BO observed preferential spreading with $R=0.4$.
More puzzeling is the large observed dispersion in the TB based calculations of the basal stacking fault and its still poor description with the MEAM potential, since it may lead to erroneous dislocations calculations in these approaches.
These results suggest that only a DFT approach or a more precise TB approach, like the DFT-based tight-binding approach (DFTB) should be able to properly describe the dislocation core structures in hcp Ti and consequently its plastic behavior.

\section{Acknowledgement}
We would like to thank Fran\c cois Willaime and Chu Chun Fu for helping us getting started with the SIESTA code. Our warm thanks go to Daniel Caillard and Alain Couret for very helpful discussions.
This work has been supported by the ANR ``SIMDIM'' contract n$^{\circ}$ BLANC-0250. This work was also granted access to the HPC resources of CALMIP under the allocation 2011-p0685. Part of the calculations have been performed on the CINES computer center.

\section*{References}

\bibliographystyle{elsarticle-num.bst}


\bibliography{biblio.bib,biblioMorillo.bib}

\end{document}